\documentclass[pre,twocolumn,amsmath,amssymb,floatfix]{revtex4-2}
\usepackage{amsthm}
\usepackage{graphicx,booktabs}
\usepackage{algpseudocode}
\newcounter{algo}
\renewcommand{\thealgo}{\arabic{algo}}
\usepackage{tikz}
\usetikzlibrary{arrows.meta,calc}
\usepackage[hidelinks]{hyperref}
\newtheorem{theorem}{Theorem}
\newtheorem{proposition}{Proposition}
\newtheorem{lemma}{Lemma}
\theoremstyle{definition}
\newtheorem{definition}{Definition}
\theoremstyle{remark}

\begin{document}
\title{Restricted triangular ice beyond the solvable manifolds: chiral ice and the critical region of dominant chiral vertices}
\author{Josep Batle}
\email{batlequantum@gmail.com}
\affiliation{Centre de Recerca Independent de sa Pobla (CRISP), 07420 sa Pobla, Balearic Islands, Spain}
\begin{abstract}
We study two restrictions of Baxter's symmetric twenty-vertex model on the triangular lattice, in which the three classes of vertices allowed by the ice rule carry the weights $a$, $b$ and $c$: chiral ice, in which only the twelve chiral vertices are allowed, and the model that excludes only the two alternating vertices. We show that neither model belongs to any of the known exactly solvable families of the twenty-vertex model, and we characterise chiral ice as the ice model in which every vertex has exactly one through line, a property that imposes a parity constraint along every closed lattice line. We compute the residual entropies of both models with transfer matrices on cylinders of up to twelve vertices in circumference, extrapolated with the finite-size correction predicted by conformal invariance, which we adapt to the triangular geometry and verify at two exactly solved points. The number of upward arrows crossing the cylinder is conserved, and the cost of changing it shows that both models are critical, with the long-distance behaviour of a free massless field. Chiral ice is the extreme point of a region of the phase diagram in which ordering had been suggested; along two paths through that region we find the same critical behaviour at every point, with no feature at the proposed boundary of the region. Along Baxter's exactly solved curve we obtain the corresponding exponents in closed form.
\end{abstract}
\maketitle

\section{Introduction}

Ice-type vertex models on the triangular lattice were introduced by Baxter, who formulated the Rys F model on this lattice and solved it by the Bethe ansatz for weights satisfying one algebraic restriction \cite{Baxter1969}. Kelland extended the solution to a four-parameter surface in the space of the ten reversal-symmetric vertex activities \cite{Kelland1974}, and Baxter's monograph discusses the integrable structure of related models \cite{BaxterBook}. The twenty-vertex model has since been studied with domain-wall boundary conditions on the integrable manifold \cite{DFG2020} and on a free-fermionic subvariety parametrised by the maximal minors of a $3\times6$ matrix \cite{Kenyon2023}. Away from these families we know of no exact results, and the phase diagram of the symmetric model has been approached through switch and loop representations \cite{Shtengel2005}.

In the symmetric model, the twenty vertices are divided into three classes invariant under the rotations and reflections of the lattice, carrying weights $a$, $b$ and $c$. Shtengel and Chayes identified a disordered region and ordered regions of the $(a,b,c)$ diagram, and left open the nature of the region $2c>b+3a$, in which chiral vertices dominate, suggesting that it might order because of the fluctuations themselves \cite{Shtengel2005}. The extreme point of that region, $(0,0,1)$, defines \emph{chiral ice}, the ice model built only from the twelve chiral vertices. The neighbouring point $(1,0,1)$, which excludes only the alternating vertices, lies inside the disordered region. Both models can be realised physically as tilings by hexagonal pieces with bumps and dents on their edges, and commercial puzzles based on this principle exist; this realisation plays no further role here.

The purpose of the present contribution is threefold. First, we establish that neither $(0,0,1)$ nor $(1,0,1)$ belongs to any of the known exactly solvable families, so that their residual entropies are new quantities. Second, we compute these entropies with controlled accuracy by transfer matrices on cylinders and a finite-size extrapolation whose amplitude is fixed by conformal invariance and by the geometry of the lattice, with no adjustable parameter, and we determine the nature of both phases from the spectrum of the transfer matrix in the sectors of fixed flux. Third, we extend the analysis to the interior of the region dominated by chiral vertices along two paths in the space of weights.

The present contribution is divided as follows. Section~\ref{sec:model} defines the model and proves the structural properties of its vertex classes, including the line characterisation of chiral ice, flux conservation and the parity constraint. Section~\ref{sec:exclusion} proves the exclusion of both points from the known solvable families. Section~\ref{sec:method} describes the transfer matrix and derives the finite-size correction for the triangular geometry. Section~\ref{sec:results} presents the residual entropies and their analysis. Section~\ref{sec:scan} examines the interior of the region dominated by chiral vertices. Finally, some conclusions are drawn in Section~\ref{sec:conclusions}. The Appendix gives the computational procedures in pseudocode.

\section{The model and its vertex classes}
\label{sec:model}

\begin{figure*}[t]
\centering
\newcommand{\tile}[7]{
\begin{scope}[shift={(#1,0)}]
 \fill[gray!55] (30:0.8)--(90:0.8)--(150:0.8)--(210:0.8)--(270:0.8)--(330:0.8)--cycle;
 \foreach \b/\a in {#2/0,#3/60,#4/120,#5/180,#6/240,#7/300}{%
   \ifnum\b=1 \fill[gray!55] (\a:0.6928) circle (0.2);
   \else \fill[white] (\a:0.6928) circle (0.2); \fi}
 \begin{scope}[shift={(0,-2.0)}]
 \foreach \b/\a in {#2/0,#3/60,#4/120,#5/180,#6/240,#7/300}{%
   \ifnum\b=1 \draw[-{Stealth[length=4pt]},thick] (0,0)--(\a:0.75);
   \else \draw[-{Stealth[length=4pt]},thick] (\a:0.75)--(\a:0.12); \fi}
 \fill (0,0) circle (1.6pt);
 \end{scope}
\end{scope}}
\begin{tikzpicture}[scale=1.15]
 \tile{0}{1}{1}{1}{0}{0}{0}
 \tile{2.8}{1}{0}{1}{0}{1}{0}
 \tile{5.6}{1}{1}{0}{1}{0}{0}
 \tile{8.4}{0}{0}{1}{0}{1}{1}
 \node at (0,-3.2) {\small consecutive (6)};
 \node at (2.8,-3.2) {\small alternating (2)};
 \node at (5.6,-3.2) {\small chiral A (6)};
 \node at (8.4,-3.2) {\small chiral B (6)};
\end{tikzpicture}
\caption{Representatives of the vertex classes of triangular ice, drawn as hexagonal tiles (a bump denotes an outgoing arrow) and as arrow configurations. Orbit sizes under rotations are given in brackets.}
\label{fig:tiles}
\end{figure*}
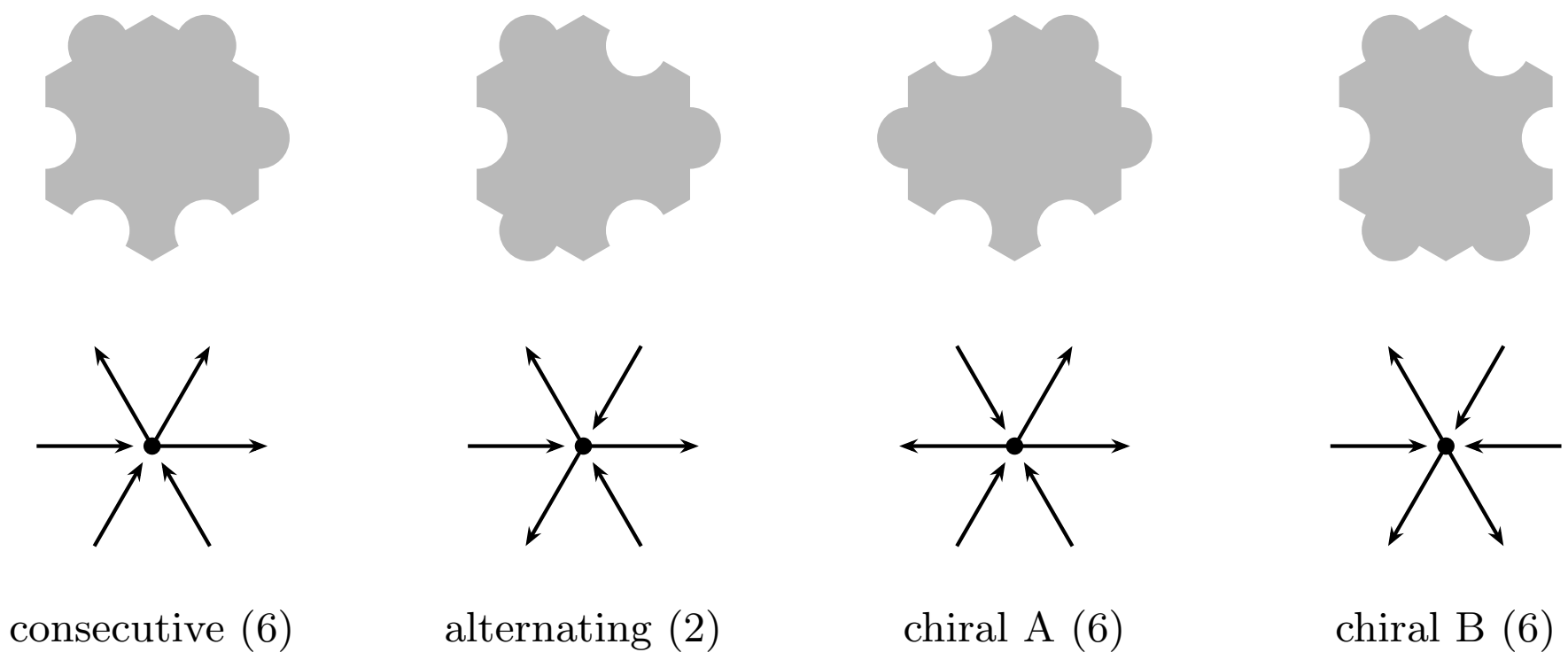

\begin{figure*}[t]
\centering
\begin{tikzpicture}[scale=1.15,>={Stealth[length=5pt]}]
 \pgfmathsetmacro{\h}{sqrt(3)/2}
 \foreach \y in {0,1,2}{ \foreach \x in {0,...,4}{ \coordinate (v\x\y) at ({\x+\y/2},{\y*\h}); } }
 \foreach \y/\arr in {0/{0,1,0,0,1},1/{0,1,0,0,0},2/{0,0,0,0,1}}{
   \foreach \x [count=\i from 1] in {0,...,3}{ \pgfmathtruncatemacro{\xx}{\x+1}
     \pgfmathtruncatemacro{\d}{{\arr}[\x]}
     \ifnum\d=1 \draw[->,gray!70!black] (v\x\y) -- (v\xx\y); \else \draw[<-,gray!70!black] (v\x\y) -- (v\xx\y); \fi }
   \draw[gray!70!black,dotted] (v4\y) -- ++(0.75,0) node[right,black,font=\scriptsize] {$x{=}1$};
   \draw[gray!70!black,dotted] (v0\y) -- ++(-0.75,0); }
 \foreach \y/\arrA/\arrB in {0/{1,0,1,1,0}/{1,0,1,0,0},1/{1,0,1,1,0}/{0,0,1,0,1}}{
   \foreach \x in {0,...,4}{
     \pgfmathtruncatemacro{\da}{{\arrA}[\x]} \pgfmathtruncatemacro{\db}{{\arrB}[\x]}
     \ifnum\da=1 \draw[->,gray!70!black] (v\x\y) -- ($(v\x\y)+(60:1)$); \else \draw[<-,gray!70!black] (v\x\y) -- ($(v\x\y)+(60:1)$); \fi
     \ifnum\db=1 \draw[->,gray!70!black] (v\x\y) -- ($(v\x\y)+(120:1)$); \else \draw[<-,gray!70!black] (v\x\y) -- ($(v\x\y)+(120:1)$); \fi } }
 \draw[green!50!black,dashed,thick] (-0.9,{0.5*\h}) -- (5.6,{0.5*\h}) node[right,font=\scriptsize] {cut $y$: $N_\uparrow=5$};
 \draw[green!50!black,dashed,thick] (-0.9,{1.5*\h}) -- (5.6,{1.5*\h}) node[right,font=\scriptsize] {cut $y{+}1$: $N_\uparrow=5$};
 \node[violet!80!black,font=\scriptsize] at ($(v20)!0.55!($(v20)+(60:1)$)+(-0.22,0.02)$) {$u_x$};
 \node[violet!80!black,font=\scriptsize] at ($(v30)!0.55!($(v30)+(120:1)$)+(0.22,0.02)$) {$v_x$};
 \draw[ultra thick,violet!60!black,-] (v20) -- ($(v20)+(60:1)$); \draw[ultra thick,violet!60!black,-] (v30) -- ($(v30)+(120:1)$);
 \draw[->,gray!70!black] (v20) -- ($(v20)+(60:1)$); \draw[<-,gray!70!black] (v30) -- ($(v30)+(120:1)$);
 \node[font=\scriptsize,above left] at (v21) {$x$};
 \node[font=\scriptsize,above] at ($(v21)!0.5!(v31)+(0,0.02)$) {$h_x$};
 \node[font=\scriptsize,above] at ($(v11)!0.5!(v21)+(0,0.02)$) {$h_{x-1}$};
 \foreach \y in {0,1,2}{ \foreach \x in {0,...,4}{ \fill (v\x\y) circle (1.5pt); } }
 \node[left,font=\scriptsize] at (-0.9,0) {row $y$};
 \node[left,font=\scriptsize] at (-0.4,{\h}) {row $y{+}1$};
 \node[left,font=\scriptsize] at (0.1,{2*\h}) {row $y{+}2$};
\end{tikzpicture}
\caption{Three consecutive rows of a cylinder of circumference $L=5$, with the ring bonds in direction $0^\circ$ and the bonds at $60^\circ$ and $120^\circ$ joining each row to the one above; the dotted segments indicate the periodic closure $x=L+1\equiv1$. The arrows form a configuration of triangular ice in the middle row. The two dashed lines are horizontal cuts, each crossing $2L=10$ bonds; the number $N_\uparrow$ of arrows pointing upwards is the same on both, as Proposition~\ref{prop:flux} states. The thick bonds are those that arrive at the vertex $x$ of row $y+1$ from below; their bits $u_x$, $v_x$ and the bits $h_{x-1}$, $h_x$ of the adjacent ring bonds are the variables of the transfer matrix of Section~\ref{sec:method}.}
\label{fig:cylinder}
\end{figure*}
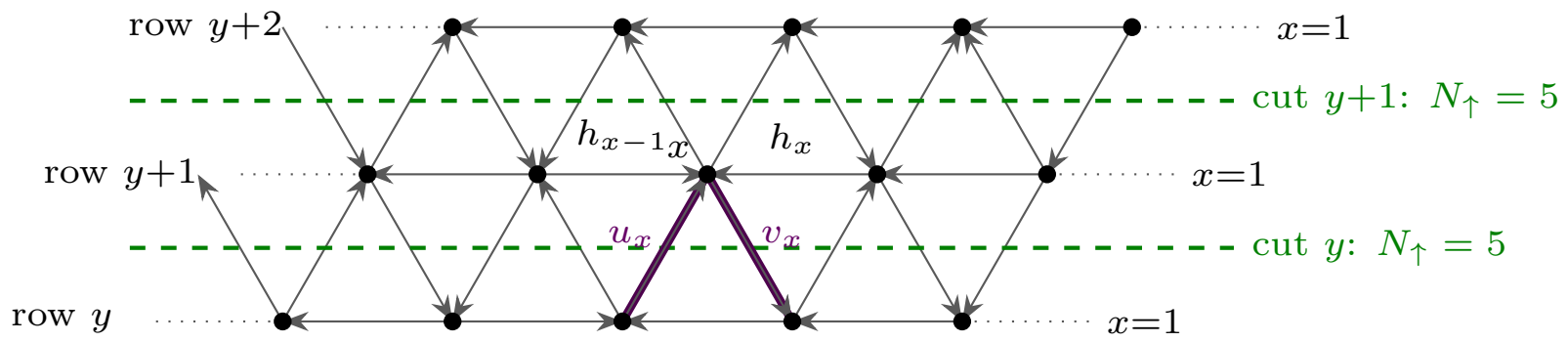

Let the triangular lattice have unit bond length, with the six neighbours of a vertex at angles $0^\circ,60^\circ,\dots,300^\circ$. An arrow is placed on every bond, and the ice rule requires three incoming and three outgoing arrows at every vertex. A vertex is described by its out-word $(\epsilon_0,\epsilon_1,\dots,\epsilon_5)\in\{0,1\}^6$, with $\epsilon_k=1$ if the arrow at angle $60k^\circ$ points outwards; the ice rule reads $\sum_k\epsilon_k=3$, which allows $\binom63=20$ vertices.

The three lattice lines through a vertex join the directions $k$ and $k+3$, $k=0,1,2$. A line is a \emph{source} at the vertex if $\epsilon_k=\epsilon_{k+3}=1$, a \emph{sink} if $\epsilon_k=\epsilon_{k+3}=0$, and a \emph{through line} otherwise.

\begin{lemma}
\label{lem:lines}
Every vertex has as many source lines as sink lines. Consequently it has either three through lines, or exactly one source line, one sink line and one through line. There are $8$ vertices of the first kind and $12$ of the second.
\end{lemma}

\begin{proof}
With $n_{\mathrm{so}}+n_{\mathrm{si}}+n_{\mathrm{th}}=3$ and $\sum_k\epsilon_k=2n_{\mathrm{so}}+n_{\mathrm{th}}=3$, subtraction gives $n_{\mathrm{so}}=n_{\mathrm{si}}$, whence $(n_{\mathrm{so}},n_{\mathrm{si}},n_{\mathrm{th}})\in\{(0,0,3),(1,1,1)\}$. In the first case each line has two orientations, giving $2^3=8$ vertices. In the second there are $3!$ assignments of the three roles to the three lines and two orientations of the through line, giving $12$ vertices.
\end{proof}

Under the dihedral group of the hexagon, the $8$ vertices with three through lines split into the orbit of $(1,1,1,0,0,0)$, of size $6$, called \emph{consecutive}, and the orbit of $(1,0,1,0,1,0)$, of size $2$, called \emph{alternating}. The $12$ vertices of the second kind, called \emph{chiral}, form one orbit under the full dihedral group and two orbits of size $6$ under rotations, represented by $(1,1,0,1,0,0)$ and $(0,0,1,0,1,1)$ and exchanged by reflections. Figure~\ref{fig:tiles} shows representatives. In Baxter's terminology the three classes are the vertices of types (i), (ii) and (iii) \cite{Baxter1969}, and in the switch analysis of Shtengel and Chayes they carry the weights $a$, $b$ and $c$ \cite{Shtengel2005}; the same letters are used here.

\begin{definition}
The \emph{symmetric triangular ice model} with weights $(a,b,c)$ assigns the weight $a$, $b$ or $c$ to every consecutive, alternating or chiral vertex respectively. \emph{Chiral ice} is the model $(0,0,1)$. The model $(1,0,1)$ is called \emph{chiral--consecutive ice}. For a model with $0/1$ weights the residual entropy per vertex is $s=\lim N^{-1}\ln W_N$, where $W_N$ is the number of configurations on $N$ vertices.
\end{definition}

The lemma yields the following characterisation, which is used repeatedly.

\begin{proposition}
\label{prop:chiral}
An ice configuration is a configuration of chiral ice if and only if every vertex has exactly one through line.
\end{proposition}

Along a lattice line, the arrows keep their direction except at the vertices where the line is a source or a sink, where the direction reverses; such vertices are called kinks of the line.

\begin{proposition}
\label{prop:parity}
On every closed lattice line the number of kinks is even. In chiral ice on a cylinder whose rows are closed lines of $L$ vertices, the number of vertices of each row at which the row is the through line has the parity of $L$.
\end{proposition}

\begin{proof}
Traversing a closed line returns the arrow direction to its initial value, and every kink reverses it, so that the number of kinks is even. In chiral ice, by Proposition~\ref{prop:chiral}, a vertex of a row is a kink of the row unless the row is its through line. The number of the latter vertices is $L$ minus the number of kinks, which has the parity of $L$.
\end{proof}

Consider now a cylinder of circumference $L$, built from rows of $L$ vertices joined by ring bonds in direction $0^\circ$, each vertex being joined to the row above by bonds in directions $60^\circ$ and $120^\circ$; Fig.~\ref{fig:cylinder} shows three consecutive rows. A horizontal cut between consecutive rows crosses the $2L$ bonds that join the two rows. Let $N_\uparrow$ be the number of these bonds whose arrow points upwards.

\begin{proposition}
\label{prop:flux}
$N_\uparrow$ is the same on every horizontal cut.
\end{proposition}

\begin{proof}
Summing $(\text{out}-\text{in})$ over the $L$ vertices of a row gives zero. Ring bonds cancel in pairs. A bond of the lower cut contributes $-1$ if it points upwards and $+1$ otherwise, and a bond of the upper cut contributes $+1$ if it points upwards and $-1$ otherwise. The sum is $2N_\uparrow^{\mathrm{upper}}-2N_\uparrow^{\mathrm{lower}}=0$.
\end{proof}

We call the integer
\begin{equation}
Q=N_\uparrow-L,\qquad -L\le Q\le L,
\label{eq:flux}
\end{equation}
the \emph{flux} through the cylinder. It is the net number of upward arrows on a cut, counted from the symmetric value $N_\uparrow=L$ at which as many arrows point up as down, and by Proposition~\ref{prop:flux} it is conserved from row to row, in the same way as a current through a tube. The flux thus labels invariant sectors of the transfer matrix defined in Section~\ref{sec:method}, and Section~\ref{sec:method} also shows how the cost of a nonzero flux distinguishes a critical phase from an ordered one. The consecutive model $(1,0,0)$ is frozen: by Lemma~\ref{lem:lines} its vertices have no kinks, every line is uniformly oriented, and the number of configurations is bounded by $2$ to the number of lines, so that $s=0$.

\section{Exclusion from the known solvable families}
\label{sec:exclusion}

\subsection{Baxter's curve}

\begin{figure*}[t]
\centering
\includegraphics[width=\textwidth]{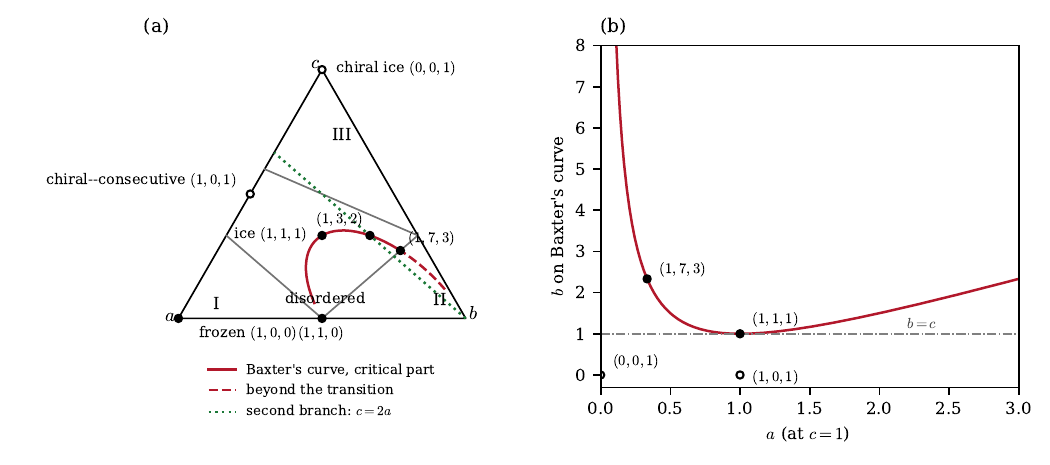}
\caption{(a) The projective plane of the weights $(a:b:c)$ in barycentric coordinates, with $a$, $b$ and $c$ at the three corners. Red: Baxter's curve \eqref{eq:baxter}, solid in its critical part between $(1:1:0)$ and $(1:7:3)$ and dashed beyond the transition. Grey: the boundaries of the ordered regions I and II and of region III of Shtengel and Chayes \cite{Shtengel2005}. Green dotted: the line $c=2a$, on which the first relation of Kelland's second branch holds in the symmetric model. The points studied in this work, chiral ice $(0,0,1)$ and chiral--consecutive ice $(1,0,1)$, are shown as open circles, and the exactly solved points used as checks as filled circles. (b) The same curve at $c=1$, where $b=1+(a-1)^2/a$: the weight $b$ never falls below $c$, which excludes $(1,0,1)$, and diverges as $a\to0$, so that the curve never reaches chiral ice.}
\label{fig:weights}
\end{figure*}

Baxter solved the symmetric model by the Bethe ansatz \cite{Baxter1969}. The method, in the form he used, seeks the eigenvectors of the row-to-row transfer matrix of Section~\ref{sec:method} in the following way. On a horizontal cut of the cylinder, a state is specified by the positions of the $n$ downward arrows among the $2L$ bonds, and the eigenvector is sought as a superposition of plane waves in these positions, in which the $n$ downward arrows play the role of particles with wave numbers $k_{1},\dots,k_{n}$ and the amplitude of each plane wave is a product of two-body scattering factors. For a generic set of weights the equations for the amplitudes are inconsistent: the three-body processes that occur when three arrows meet at a vertex do not reduce to products of two-body ones. Baxter found that they do reduce exactly when the weights satisfy \cite[eq.~(22)]{Baxter1969}
\begin{equation}
(a-c)^2=a(b-c).
\label{eq:baxter}
\end{equation}
On this curve the periodicity of the wave function around the cylinder gives a set of $n$ coupled algebraic equations for the wave numbers, the Bethe equations, and in the limit of large $L$ their solutions condense into a density on the real axis that satisfies a linear integral equation; the free energy follows from this density, and so does the finite-size spectrum, as Section~\ref{sec:method} shows. Figure~\ref{fig:weights}(a) locates the curve in the plane of the weights $(a:b:c)$, which only matter up to a common factor. The curve contains the ice point $a=b=c$, the point $b=a$, $c=0$, and the infinite-order transition at $b=7a$, $c=3a$; its physical part, on which the model is critical, runs from $(1:1:0)$ to $(1:7:3)$, and beyond that point the model is in an ordered antiferroelectric phase.

\begin{theorem}
\label{thm:baxter}
No point of \eqref{eq:baxter} with $a>0$ satisfies $b<c$. The points $(1,0,1)$ and $(0,0,1)$ do not lie on \eqref{eq:baxter}, and no sequence of points on \eqref{eq:baxter} with $c$ bounded away from zero converges projectively to $(0,0,1)$.
\end{theorem}

\begin{proof}
For $a>0$, \eqref{eq:baxter} gives $b-c=(a-c)^2/a\ge0$. The point $(1,0,1)$ has $b-c=-1$. At $a=0$, \eqref{eq:baxter} reduces to $c^2=0$, which excludes $(0,0,1)$. Along the curve with $a\to0$ and $c\ge c_0>0$, $b=c+(a-c)^2/a\ge c_0^2/a-c$ diverges, so that the projective limit is $(0:1:0)$.
\end{proof}
This exclusion is analytic and involves no numerical input.

\subsection{Kelland's surface}

Kelland extended Baxter's solution to a model in which the twenty vertices carry ten independent weights, each vertex having the same weight as the vertex obtained by reversing all of its arrows \cite{Kelland1974}. He calls these weights activities, $u_1,\dots,u_{10}$, and the symmetric model is the special case in which they take only the three values $a$, $b$ and $c$. His construction is again a Bethe ansatz, in which an eigenvector is built from states with a given number of reversed lines; the ansatz with a single reversed line, which he calls the one-magnon problem by analogy with the spin-wave excitations of a ferromagnet, imposes algebraic conditions on the activities, the commutation conditions, and the full ansatz works on the four-parameter surface $S$ of the ten-dimensional space of activities on which these conditions hold.

To identify Kelland's activities with the vertex classes of Section~\ref{sec:model} we use two of his statements. Under a rotation by $\pi/3$ the activities are permuted as $(1\,2\,3)(4)(5\,7\,9\,6\,8\,10)$ \cite[eq.~(34)]{Kelland1974}, so that $u_4$ is the activity of the orbit of size $2$, $u_1,u_2,u_3$ those of an orbit of size $6$ that is fixed by the rotation by $2\pi/3$, and $u_5,\dots,u_{10}$ those of the remaining twelve vertices. Secondly, the free energy of the frozen ferroelectric states is $\ln\max(u_1,u_2,u_3)$ \cite[eq.~(65)]{Kelland1974}. The name comes from the origin of ice-type models as models of hydrogen-bonded ferroelectrics such as potassium dihydrogen phosphate \cite{Slater1941,BaxterBook}, in which an arrow stands for the position of a proton on a bond and hence for an electric dipole. A state in which every lattice line is uniformly oriented is frozen, because reversing any arrow would create a kink on its line and, by the parity argument of Proposition~\ref{prop:parity}, a second one, so that no local change is possible; in such a state every vertex has three through lines, and by Lemma~\ref{lem:lines} it is consecutive or alternating. The ferroelectric frozen states are those with a net polarisation, in which the three orientations at a vertex add to a dipole, and these are built from consecutive vertices, whose outgoing arrows occupy one half of the hexagon; the alternating vertex has no dipole. The free energy $\ln\max(u_1,u_2,u_3)$ of these states therefore identifies $u_1,u_2,u_3$ as the activities of the consecutive class. Hence $u_4$ is the activity of the alternating pair and $u_5,\dots,u_{10}$ those of the chiral vertices, and the symmetric model corresponds to $u_1=u_2=u_3=a$, $u_4=b$, $u_5=\dots=u_{10}=c$. Kelland's surface $S$ is defined by \cite[eqs.~(19)]{Kelland1974}
\begin{align}
&u_5=u_6,\quad u_7=u_8,\quad u_9=u_{10},\label{eq:k19abc}\\
&I=u_2u_3+\frac{u_1u_5u_9}{u_7}=u_3u_1+\frac{u_2u_7u_5}{u_9}=u_1u_2+\frac{u_3u_7u_9}{u_5},\label{eq:k19d}\\
&u_4=\frac{I^2}{u_1u_2u_3}-\left(\frac1{u_1}+\frac1{u_2}+\frac1{u_3}\right)I+u_1+u_2+u_3.\label{eq:k19e}
\end{align}

\begin{proposition}
\label{prop:slice}
The intersection of $S$ with the symmetric model is Baxter's curve \eqref{eq:baxter}, the red curve of Fig.~\ref{fig:weights}(a).
\end{proposition}

\begin{proof}
With $u_1=u_2=u_3=a$ and $u_5=\dots=u_{10}=c$, all three expressions in \eqref{eq:k19d} equal $I=a^2+ac$. Then \eqref{eq:k19e} gives
\begin{equation*}
\begin{aligned}
b&=\frac{(a^2+ac)^2}{a^3}-\frac{3(a^2+ac)}{a}+3a\\
&=\frac{(a+c)^2}{a}-3c=\frac{a^2-ac+c^2}{a},
\end{aligned}
\end{equation*}
which is equivalent to $(a-c)^2=a(b-c)$.
\end{proof}

Since Kelland's model has no external field, the activities of a vertex and of its reversal coincide and no gauge transformation relates distinct points of the $u$-space. Chiral--consecutive ice therefore corresponds to the single point $u_1=u_2=u_3=1$, $u_4=0$, $u_5=\dots=u_{10}=1$, and chiral ice to $u_1=u_2=u_3=u_4=0$, $u_5=\dots=u_{10}=1$.

\begin{theorem}
\label{thm:kelland}
Chiral--consecutive ice does not lie on $S$, and it violates the one-magnon commutation conditions on which both branches of Kelland's ansatz rest. Chiral ice does not lie on $S$ at any finite value of Kelland's parameters, it violates the unstudied branch of his commutation conditions, and it does not belong to the Zariski closure of $S$.
\end{theorem}

\begin{proof}
For chiral--consecutive ice, Proposition~\ref{prop:slice} requires $b=(a^2-ac+c^2)/a=1\ne0$. Kelland's commutation conditions are expressed through the quantities $v_1=u_9u_{10}/(u_1u_2)$, $v_2=u_7u_8/u_1^2$, $w_1=u_6u_{10}/(u_1u_7)$, $w_2=u_2u_3/u_1^2$, $w_3=u_2u_5u_7/(u_1^2u_{10})$, $w_4=u_4/u_1$, $w_5=u_5u_{10}/(u_2u_7)$, $w_6=u_3/u_2$ and $w_7=u_6u_7/(u_1u_{10})$ \cite[eq.~(A12)]{Kelland1974}. At chiral--consecutive ice all of them equal $1$ except $w_4=0$. The condition $w_3(w_4-v_2)=w_2(w_7-v_2)$ \cite[eq.~(A13c)]{Kelland1974} then reads $-1=0$. On the symmetric model the same condition reduces to $c(b-c^2)=c-c^2$ in units $a=1$, which is \eqref{eq:baxter}.

For chiral ice we use Kelland's explicit parametrisation of $S$ \cite[eqs.~(33)]{Kelland1974}: the surface is the image of three angles $\theta$, $\phi$ and $\zeta$, together with an overall scale, and the activities relevant here are, up to that common factor,
\begin{gather*}
u_1\propto\sin(\theta+\zeta)\sin(\phi-\theta+\zeta)\sin2\theta,\\
u_2\propto\sin(\theta-\zeta)\sin(\phi-\theta-\zeta)\sin2\theta,\\
u_5\propto\sin(\phi-\theta+\zeta)\sin(\phi-\theta-\zeta)\sin\phi,\\
u_7\propto\sin(\phi-\theta+\zeta)\sin\phi\sin2\theta.
\end{gather*}
Chiral ice requires $u_1=u_2=0$ with $u_5,u_7\ne0$. The conditions $u_5\ne0$ and $u_7\ne0$ require $\sin(\phi-\theta\pm\zeta)\ne0$ and $\sin2\theta\ne0$, so that $u_1=u_2=0$ forces $\sin(\theta+\zeta)=\sin(\theta-\zeta)=0$. Since the zeros of the sine are the real multiples of $\pi$, also for complex arguments, this gives $2\theta\in\pi\mathbb Z$ and $\sin2\theta=0$, a contradiction: no values of the angles, real or complex, reach chiral ice. Figure~\ref{fig:kelland} shows the argument for real angles: the lines on which $u_1$ and $u_2$ vanish cross only on the lines where $u_7$ vanishes as well.

\begin{figure}[t]
\centering
\includegraphics[width=\columnwidth]{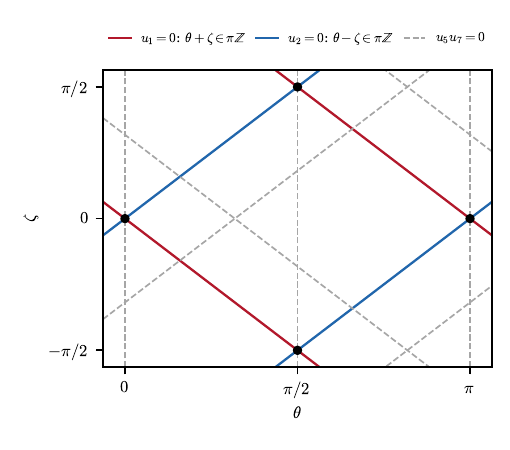}
\caption{Kelland's parametrisation near chiral ice, in the plane of the angles $(\theta,\zeta)$ at $\phi=1$. Red: $u_1=0$ with $u_5u_7\ne0$, that is, $\theta+\zeta\in\pi\mathbb Z$; blue: $u_2=0$ with $u_5u_7\ne0$, that is, $\theta-\zeta\in\pi\mathbb Z$; grey dashed: the lines on which $u_5$ or $u_7$ vanishes. Chiral ice requires a point on a red and a blue line but on no grey line; every crossing of a red and a blue line (dots) lies on a grey line $\sin2\theta=0$.}
\label{fig:kelland}
\end{figure}

The commutation conditions admit two solutions, which Kelland calls branches; he develops the first and leaves the second unstudied. The second branch is $v_2=w_3+w_7$, $v_1=w_1+w_5$ \cite[eq.~(A14)]{Kelland1974}. Clearing denominators, the first relation becomes $u_7u_8u_{10}=u_2u_5u_7+u_1u_6u_7$, which at chiral ice reads $1=0$. At chiral--consecutive ice the same relation reads $1=2$. In the symmetric model the relation becomes $c^{3}=2ac^{2}$, that is, $c=2a$: the dotted line of Fig.~\ref{fig:weights}(a), which passes through neither of the two points.

It remains to exclude chiral ice from the closure of $S$, that is, from the set of limits of points of $S$. The cleared equations \eqref{eq:k19abc}--\eqref{eq:k19e} are satisfied trivially at chiral ice, because \eqref{eq:k19e} becomes $0/0$. The genuine surface requires $u_1u_2u_3u_5u_7u_9\ne0$, since Kelland divides by these activities. Its closure in the sense of algebraic geometry, the Zariski closure, is the set of common zeros of all polynomials that vanish on $S$. These polynomials form an ideal, which is obtained from the ideal of the cleared equations by two standard operations: saturation with respect to the product above, which discards the spurious solutions on which some of these activities vanish, and elimination of the auxiliary variable $I$. Both operations are carried out with a Gr\"obner basis, a set of generators of the ideal from which variables can be eliminated in a prescribed order, much as Gaussian elimination does for linear equations. This elimination, given in pseudocode as Algorithm~\ref{alg:closure} of the Appendix, shows that the generators of this ideal vanish at the ice point $(1,1,1)$, at the point $(1,3,2)$ and at the generic point $(1,7/9,1/3)$ of Baxter's curve, as they must, and that they do not all vanish at chiral ice or at chiral--consecutive ice.
\end{proof}

The mechanism behind the closure statement can be seen at leading order, and Fig.~\ref{fig:weights}(b) shows it on the symmetric slice. Near chiral ice, \eqref{eq:k19d} gives $I\simeq u_1u_5u_9/u_7\simeq u_2u_5u_7/u_9\simeq u_3u_7u_9/u_5\equiv J$, and \eqref{eq:k19e} then gives $u_4\simeq u_5u_7u_9/J$. Approaching vanishing consecutive activities on $S$ therefore drives the alternating activity to infinity, exactly as Baxter's formula does on the symmetric slice, where $b=1+(a-1)^{2}/a$ at $c=1$ diverges as $a\to0$. In Theorem~\ref{thm:kelland}, the exclusion of chiral--consecutive ice, the violation of the commutation conditions and the impossibility at finite parameters are analytic; the exclusion from the Zariski closure is computer-assisted, resting on the elimination of Algorithm~\ref{alg:closure}.

Di Francesco and Guitter use integrable weights which they attribute to Kelland \cite{Kelland1974} and to Baxter, written in a four-parameter projective form and built from six-vertex weights on the kagome lattice \cite{DFG2020}; Theorem~\ref{thm:kelland} applies to them insofar as they parametrise Kelland's surface. Kelland notes that other surfaces on which a Bethe ansatz works almost certainly exist, citing as an example the analogue of the free-fermion case \cite{Kelland1974}; this is the family we treat next.

\subsection{The free-fermionic subvariety}

A vertex model is called free-fermionic when its partition function can be written as a Pfaffian or a determinant, so that it describes non-interacting fermions. On the square lattice this is the case for the six-vertex model when the weights satisfy the free-fermion condition $\omega_1\omega_2+\omega_3\omega_4=\omega_5\omega_6$ of Fan and Wu \cite{FanWu1970}: the transfer matrix is then a quadratic form in fermionic creation and annihilation operators, which a Fourier transform diagonalises, exactly as for a lattice of dimers. For the twenty-vertex model the free-fermionic weights were identified by Kenyon and Ovenhouse, who showed that the model is then a dimer model of higher rank \cite{Kenyon2023}. They parametrise the free-fermionic twenty-vertex weights as the maximal minors $\Delta_{ijk}$ of a $3\times6$ matrix whose columns correspond to the six edges at a vertex in cyclic order, the weight of a vertex being the minor indexed by its three edges of a given orientation \cite{Kenyon2023}. With this labelling, the alternating vertices correspond to $\{1,3,5\}$ and $\{2,4,6\}$, the consecutive vertices to the six cyclic intervals, and the chiral vertices to the remaining twelve triples. A column rescaling multiplies each minor by a product of column factors and contains, in particular, the field-type gauge transformations, which leave the partition function invariant within a flux sector.

\begin{theorem}
\label{thm:ff}
Neither chiral ice nor chiral--consecutive ice belongs to the free-fermionic subvariety, even up to column rescaling.
\end{theorem}

\begin{proof}
The maximal minors of a matrix are not independent: they satisfy quadratic identities, the Pl\"ucker relations, which hold for every matrix because the minors are the coordinates of the subspace spanned by its rows. Every $3\times6$ matrix satisfies, in particular, the three-term Pl\"ucker relation
\begin{equation}
\Delta_{124}\Delta_{135}=\Delta_{123}\Delta_{145}+\Delta_{125}\Delta_{134}.
\label{eq:plucker}
\end{equation}
Here $\{1,3,5\}$ is alternating, $\{1,2,3\}$ is consecutive, and $\{1,2,4\}$, $\{1,4,5\}$, $\{1,2,5\}$ and $\{1,3,4\}$ are chiral, as follows from Lemma~\ref{lem:lines} by inspecting their lines $\{1,4\}$, $\{2,5\}$, $\{3,6\}$. For chiral ice, after any column rescaling, the left-hand side of \eqref{eq:plucker} and the first term on the right vanish while the second term is a product of two nonzero chiral weights, a contradiction. For chiral--consecutive ice the left-hand side vanishes while both terms on the right are positive.
\end{proof}

The proof of Theorem~\ref{thm:ff} is analytic. Theorems~\ref{thm:baxter}--\ref{thm:ff} establish that both points lie outside every exactly solvable family of the twenty-vertex model known to us. They do not exclude a solvable structure that has not yet been found.

\subsection{Position of the two models in the phase diagram}

Having placed the two models outside the solvable families, we now place them in the phase diagram of the symmetric model, Fig.~\ref{fig:weights}(a), in order to state what question about that diagram their study answers. Shtengel and Chayes rewrote the ice-rule models in terms of \emph{switches}: a switch at a vertex is a pairing of its six bonds into three pairs, each pair being traversed by a loop that enters along one bond and leaves along the other, and a vertex weight is the sum of the weights of the switches consistent with the arrows at that vertex. Summing over the orientations of the loops gives back the vertex model, and the positivity of the switch weights gives the loop representation a probabilistic meaning, on which their argument for a disordered phase rests. In the symmetric model the five classes of switches have weights $\alpha,\dots,\varepsilon$ and the three vertex weights are linear in them, $a+2c=b+8\gamma+4\delta$, $b+2c=a+4(\alpha+\beta+\gamma)$ and $b+3a=2c+4(\beta+\delta+\varepsilon)$ \cite{Shtengel2005}, so that positive switch weights exist exactly when $a+2c>b$, $b+2c>a$ and $b+3a>2c$. They found with ordered regions I ($a>b+2c$) and II ($b>a+2c$) and a third region $2c>b+3a$ whose nature they left open: there the loop representation fails because some ground states have local zero modes, and they suggested an entropically stabilised phase, that is, ordering by disorder \cite{Shtengel2005}, the mechanism first described by Villain and coworkers \cite{Villain1980} by which fluctuations, rather than the energy, select an ordered state among many equivalent ones. Chiral--consecutive ice satisfies the three inequalities and lies in the disordered region. Chiral ice satisfies the first two but violates the third, and lies at the extreme point of region III. Both endpoints of the physical part of Baxter's curve, $(1:1:0)$ and $(1:7:3)$, lie on the boundary $b=a+2c$ of region II.

\section{Transfer matrix and finite-size correction}
\label{sec:method}

The partition function on a cylinder is computed row by row with a transfer matrix, which we now construct; Fig.~\ref{fig:cylinder} shows the geometry and the variables. Label the rows of the cylinder by $y$ and the vertices of a row by $x=1,\dots,L$, with $x+L\equiv x$, so that vertex $(x,y)$ is joined to $(x+1,y)$ by a ring bond, to $(x,y+1)$ by the bond at $60^\circ$ and to $(x-1,y+1)$ by the bond at $120^\circ$. The $2L$ bonds crossing the cut between rows $y$ and $y+1$ are described by $2L$ bits, and the state of the cut is $\sigma=(u_1,v_1,\dots,u_L,v_L)$, where $u_x$ and $v_x$ are the bits of the two bonds that arrive at vertex $x$ of row $y+1$ from below, from $(x,y)$ at $60^\circ$ and from $(x+1,y)$ at $120^\circ$, each equal to $1$ when the arrow points upwards; a cut thus has $2^{2L}$ states. The ring bonds of row $y+1$ carry the bits $h_x$, equal to $1$ when the arrow from $x$ to $x+1$ points to the right. Once the state $\sigma$ of the cut below, the state $\sigma'$ of the cut above and the ring bits are given, every arrow around vertex $x$ of row $y+1$ is known: in the out-word convention of Section~\ref{sec:model}, at the angles $0^\circ,60^\circ,\dots,300^\circ$, the vertex sees the ring bond to $x+1$, its two bonds to the row above, the ring bond from $x-1$, and its two bonds from the row below, so that
\begin{equation}
\varepsilon(x)=\bigl(h_x,\ u'_x,\ v'_{x-1},\ 1-h_{x-1},\ 1-u_x,\ 1-v_x\bigr),
\label{eq:word}
\end{equation}
and
\begin{equation}
\begin{aligned}
T_{\sigma'\sigma}&=\sum_{h_1,\dots,h_L\in\{0,1\}}\ \prod_{x=1}^{L}w\bigl(\varepsilon(x)\bigr),\\
h_0&\equiv h_L,\qquad v'_0\equiv v'_L,
\end{aligned}
\label{eq:Tdef}
\end{equation}
where $w(\varepsilon)$ equals $a$, $b$ or $c$ according to the class of the vertex and vanishes if $\varepsilon$ does not contain three ones, that is, if the ice rule fails. The sum over the ring bits, with the closure $h_0=h_L$ that joins the last vertex of the row to the first, runs over the $2^{L}$ ways of completing the row between the two cut states; $T_{\sigma'\sigma}$ is the total weight of these completions, and for weights in $\{0,1\}$ it counts the admissible ones. For $a=b=c=1$ this is the transfer matrix of unweighted triangular ice. On a torus of $M$ rows the two cut states at the top and the bottom are identified, and the sum over all configurations factorises row by row into $Z=\operatorname{Tr}T^M$; for $M\to\infty$ the trace is dominated by the largest eigenvalue $\lambda_{\max}(T)$, so that the entropy per vertex of a cylinder of circumference $L$ is $s(L)=L^{-1}\ln\lambda_{\max}(T)$. By Proposition~\ref{prop:flux}, $T$ connects only states with the same $N_\uparrow$, so that it is block diagonal in the flux \eqref{eq:flux}, with blocks of dimension $\binom{2L}{N_\uparrow}$; arrow reversal maps every model considered onto itself and $N_\uparrow$ onto $2L-N_\uparrow$, so that the blocks $Q$ and $-Q$ have the same spectrum. The numerical implementation applies \eqref{eq:Tdef} vertex by vertex, as a sequence of contractions with the six-index tensor of vertex weights (Algorithm~\ref{alg:T} of the Appendix). It was checked by comparing $\operatorname{Tr}T^M$ with a direct enumeration of all $2^{3LM}$ orientations on the tori $(L,M)=(2,2),(2,3),(3,2)$ for the four models $(1,1,1)$, $(1,0,1)$, $(0,0,1)$ and $(1,0,0)$.

For a critical model whose continuum limit is isotropic, the free energy per unit length of a cylinder of circumference $W$ satisfies \cite{Blote1986,Affleck1986}
\begin{equation}
\frac{\ln Z}{\text{length}}=gW+\frac{\pi c}{6W}+o(W^{-1}),
\label{eq:cft}
\end{equation}
where $g$ is the bulk free energy per unit area, with the sign convention $\ln Z=g\times\text{area}$, and $c$ is the central charge. With unit bonds, $W=L$ and the row spacing is $\sqrt3/2$, so that $\ln\lambda_{\max}=\frac{\sqrt3}{2}\left(gL+\frac{\pi c}{6L}\right)$ and
\begin{equation}
s(L)=s_\infty+\frac{\sqrt3\,\pi c}{12\,L^2}+o(L^{-2}),\qquad s_\infty=\frac{\sqrt3}{2}g.
\label{eq:fss}
\end{equation}
The isotropy required by \eqref{eq:cft} in the Euclidean metric of the regular embedding follows from the sixfold rotational symmetry of all models considered, since a symmetric second-rank tensor invariant under rotations by $\pi/3$ is a multiple of the identity.

\begin{figure*}[t]
\centering
\includegraphics[width=\textwidth]{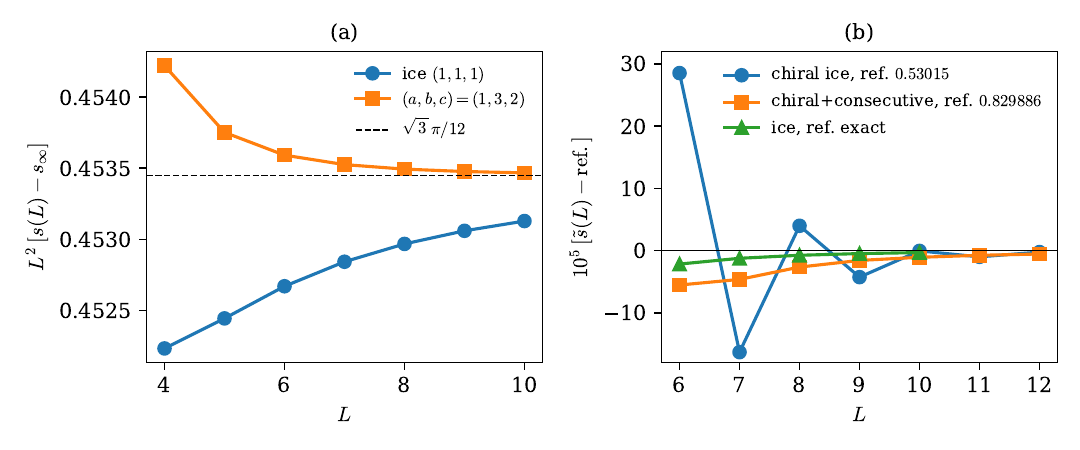}
\caption{(a) $L^2[s(L)-s_\infty]$ at the exactly solved points $(1,1,1)$ and $(1,3,2)$ compared with the parameter-free amplitude $\sqrt3\pi/12$. (b) Corrected estimates $\tilde s(L)=s(L)-\sqrt3\pi/(12L^2)$ for chiral ice, chiral--consecutive ice and the ice point, measured from the indicated references.}
\label{fig:scaling}
\end{figure*}

The amplitude in \eqref{eq:fss} was tested at two exactly solved points of Baxter's curve. At $(1,1,1)$, $s_\infty=\ln(3\sqrt3/2)=0.9547713$ \cite{Baxter1969}. At $(1,3,2)$, Baxter's case $w^3=-1$ gives \cite[eq.~(42)]{Baxter1969}
\begin{equation}
s_\infty=\frac{3\sqrt3}{\pi}\sum_{n\ \mathrm{odd}}\frac{\chi_{-3}(n)}{n^2}=\frac{15\sqrt3}{4\pi}\,L(2,\chi_{-3})=1.6153297,
\label{eq:132}
\end{equation}
where $\chi_{-3}(n)$ equals $1$, $-1$ or $0$ according as $n\equiv1$, $2$ or $0$ modulo $3$, so that the series reads $1-\tfrac1{25}+\tfrac1{49}-\tfrac1{121}+\tfrac1{169}-\dots$, $L(2,\chi_{-3})=\sum_{n\ge1}\chi_{-3}(n)/n^{2}=0.7813024$ is the corresponding Dirichlet $L$-function, and the sum over odd $n$ equals
\begin{equation*}
\Bigl(1-\tfrac14\chi_{-3}(2)\Bigr)L(2,\chi_{-3})=\tfrac54L(2,\chi_{-3}).
\end{equation*} The point $(1,3,2)$ is also the circularly symmetric free-fermionic point of \cite{Kenyon2023}, whose minors give molecule weights in the ratio $1:2:3$ for consecutive, chiral and alternating vertices. Figure~\ref{fig:scaling}(a) shows that $L^2[s(L)-s_\infty]$ converges to $\sqrt3\pi/12=0.453450$ at both points, from below at $(1,1,1)$ and from above at $(1,3,2)$; at $L=8$ the values are $0.452969$ and $0.453494$. This confirms \eqref{eq:fss} with $c=1$ and validates the geometric factor and the transfer-matrix implementation simultaneously. The amplitude has been followed to $L=10$ at both points, where $L^2[s(L)-s_\infty]$ equals $0.453129$ and $0.453468$.

The flux sectors give access to scaling dimensions. Recall from Section~\ref{sec:model} that the flux $Q=N_\uparrow-L$ of Eq.~\eqref{eq:flux} is conserved from row to row, so that $T$ can be diagonalised in each sector separately; let $\Lambda_Q$ denote its largest eigenvalue in the sector of flux $Q$. The transfer matrix can be read as an evolution operator in imaginary time: writing $T=\exp(-\tau H)$ with $\tau=\sqrt3/2$ the spacing between rows, $H$ is the Hamiltonian of a quantum chain whose sites are the $2L$ bonds of a cut, that is, of a system living on the circle of circumference $L$ that forms the cross-section of the cylinder, and the eigenvalues of $T$ are the Boltzmann factors $e^{-\tau E}$ of its energy levels. For a critical system the conformal map from the plane to the cylinder relates the energy gap between the ground state and the lowest state carrying an operator of scaling dimension $x$ to the circumference, $\Delta E=2\pi x/L$ \cite{Cardy1984}. The lowest state of sector $Q$ therefore has
\begin{equation}
x_Q(L)=\frac{L}{\sqrt3\,\pi}\ln\frac{\Lambda_0}{\Lambda_Q}.
\label{eq:xQ}
\end{equation}
The physical meaning of these dimensions rests on the height representation of ice models. Because three arrows enter and three leave every vertex, one can assign to the faces of the lattice an integer height that increases by one when a bond is crossed with its arrow pointing to the right and decreases by one otherwise; the ice rule makes the height single-valued on any simply connected region. On the cylinder the height changes by $Q$ around the circumference, so that the flux is the winding of the height. In a critical phase the coarse-grained height fluctuates with the free energy of a free massless field, $\frac{K}{2}\int(\nabla h)^{2}$, called Gaussian or Coulomb phase, and a winding $Q$ around a cylinder of circumference $L$ costs a free energy proportional to $Q^{2}/L$: the flux sectors then carry dimensions $x_Q=\kappa Q^2$, with a model-dependent constant $\kappa$ proportional to the stiffness $K$, so that $x_2/x_1\to4$. In an ordered phase the height is locked to a flat configuration, a winding must be accommodated by a domain wall along the cylinder, and the cost of a nonzero flux is a free energy per row that does not vanish as $L\to\infty$, so that $x_Q(L)$ as defined by \eqref{eq:xQ} grows linearly with $L$. The two behaviours are therefore easily told apart at the circumferences available. The sector eigenvalues were obtained with the implicitly restarted Arnoldi method, an iterative procedure that computes a few extreme eigenvalues of a large matrix from repeated applications of the matrix to vectors, here the vectors of a single sector, without ever storing the matrix; the application of $T$ was checked against an independent implementation, and no component outside the starting sector was generated, as Proposition~\ref{prop:flux} requires.

The quantity relevant for the residual entropy is the largest eigenvalue of $T$ over all sectors, whereas the values reported below are those of the sector $Q=0$. For $L\le9$ the largest eigenvalue of $T$ on the full state space, obtained by power iteration from a strictly positive vector, coincides with $\Lambda_0$ to ten significant digits for all the models considered. For $10\le L\le12$ the same comparison, carried out with the Arnoldi method on the full state space for chiral and chiral--consecutive ice, gives $\Lambda_{\mathrm{full}}/\Lambda_0=1$ to within $6\times10^{-10}$, the precision of the stored reference values. Together with the arrow-reversal symmetry $\Lambda_Q=\Lambda_{-Q}$, this establishes that the zero-flux sector carries the leading eigenvalue at all circumferences used.

On Baxter's curve the flux dimensions can be obtained in closed form from the Bethe-ansatz equations of \cite{Baxter1969}. For $0<c<3a$ Baxter introduces $w=\mathrm{e}^{-\mathrm{i}\phi}$ with $c/a=1+w+w^{-1}$, that is
\begin{equation}
\frac ca=1+2\cos\phi,\qquad 0<\phi<\frac{2\pi}{3}.
\label{eq:phi}
\end{equation}
\begin{proposition}
\label{prop:dressed}
For weights on Baxter's curve \eqref{eq:baxter} with $0<c<3a$, the lowest state of the sector of flux $Q$ has scaling dimension
\begin{equation}
x_Q=\frac{2\pi-3\phi}{4\pi}\,Q^2 .
\label{eq:xQexact}
\end{equation}
\end{proposition}
\begin{proof}[Derivation]
We use Baxter's Bethe equations for a cylinder of circumference $N$ with $n$ downward arrows per cut \cite[eqs.~(35)--(36)]{Baxter1969}. Taking their logarithm and parametrising the wave numbers by rapidities $u_j$, they read
\begin{equation}
N\bigl[F(\phi,u_j)+F(2\phi,u_j)\bigr]=I_j+\sum_{l=1}^{n}F(3\phi,u_j-u_l),
\label{eq:bethe}
\end{equation}
for $j=1,\dots,n$, where the $I_j$ are distinct numbers, integers or half-odd integers according to the parity of $n$, that label the state, and
\begin{equation}
F(\theta,u)=\frac1\pi\arctan\Bigl[\cot\frac\theta2\tanh\frac u2\Bigr]
\label{eq:Fdef}
\end{equation}
is the phase function; the two terms on the left are the momentum of a root, and the term on the right is the phase shift of the two-body scattering. The ground state of the sector $Q=0$ has $n=N$ roots with consecutive $I_j$. In the limit $N\to\infty$ at fixed $n/N$ the roots become dense, with a density $\rho(u)$ such that $N\rho(u)\,\mathrm du$ roots lie in $\mathrm du$; dividing \eqref{eq:bethe} by $N$ and differentiating with respect to $u_j$, with $\mathrm d(I_j/N)/\mathrm du_j=\rho(u_j)$ and the sum replaced by an integral, gives the linear integral equation
\begin{equation}
\rho(u)+\int_{-\infty}^{\infty}K(u-u')\rho(u')\,\mathrm du'=\partial_u\bigl[F(\phi,u)+F(2\phi,u)\bigr],
\label{eq:lieb}
\end{equation}
with the kernel
\begin{equation}
K(u)=\partial_uF(3\phi,u)=\frac{1}{2\pi}\,\frac{\sin3\phi}{\cosh u-\cos3\phi},
\label{eq:kernel}
\end{equation}
obtained from \eqref{eq:Fdef} by the identity $\sin^{2}\frac\theta2\cosh^{2}\frac u2+\cos^{2}\frac\theta2\sinh^{2}\frac u2=\frac12(\cosh u-\cos\theta)$. For $n=N$ the roots fill the whole real axis. Using $\int_{-\infty}^{\infty}\mathrm du/(\cosh u-\cos\theta)=2(\pi-\theta)/\sin\theta$ for $0<\theta<2\pi$, the integral of the kernel is $\hat K(0)=\int K\,\mathrm du=(\pi-3\phi)/\pi$.

The finite-size spectrum of a Bethe-ansatz solvable model with a single species of roots is governed by the dressed charge $Z$ \cite{KBI1993}. It is defined as the solution of
\begin{equation}
Z(u)+\int_{-\Lambda}^{\Lambda}K(u-u')Z(u')\,\mathrm du'=1
\label{eq:dressed}
\end{equation}
over the interval $[-\Lambda,\Lambda]$ occupied by the roots, evaluated at its boundary, $Z=Z(\Lambda)$, and it measures the change of the total number of roots when one root is added at the boundary of the distribution, including the rearrangement of the others; the gaps of the Hamiltonian are those of a free boson whose compactification radius is set by $Z$, and a state with $\Delta N$ roots added to the ground state has dimension $x=\Delta N^2/(4Z^2)$ \cite{KBI1993}. Only the kernel enters this equation: the result depends on the two-body phase shift $F(3\phi,u)$ alone, the momentum terms of \eqref{eq:bethe} serving only to fix the ground-state distribution of the roots, which fills the whole real axis for $n=N$. Here $\Lambda\to\infty$. In this limit the boundary value $Z(\Lambda)$ is not the constant $1/[1+\hat K(0)]$ that solves \eqref{eq:dressed} on the whole line, because the solution near the edge of the interval differs from the solution in its interior however large $\Lambda$ is. The limit is obtained by the Wiener--Hopf method, which solves the integral equation on a half-line by factorising $1+\hat K(k)$ into functions analytic in the upper and lower halves of the complex $k$-plane, and gives $Z(\Lambda)^{2}\to1/[1+\hat K(0)]$ as $\Lambda\to\infty$ \cite{KBI1993}; hence $Z^2=\pi/(2\pi-3\phi)$. Each root is a downward arrow on a horizontal cut, so that adding $\Delta N$ roots changes the flux by $Q=-\Delta N$, and \eqref{eq:xQexact} follows.
\end{proof}
At the ice point, $\phi=\pi/2$ and \eqref{eq:xQexact} gives $x_1=1/8$; at $(1,3,2)$, $\phi=\pi/3$ and $x_1=1/4$, the free-fermion value, as befits the free-fermionic point of \cite{Kenyon2023}. The transfer-matrix values agree with both: $x_1(10)=0.12496$ at the ice point, and $x_1(8)=0.25002$, $x_2(8)/x_1(8)=4.0000$ at $(1,3,2)$. These two checks validate \eqref{eq:xQ}, including its geometric factor, independently of the central charge. The dimension \eqref{eq:xQexact} coincides with that of the lowest magnetic excitation of a six-vertex model of anisotropy angle $\gamma=3\phi/2$, namely $(\pi-\gamma)/(2\pi)$, in line with the construction of the integrable twenty-vertex weights from six-vertex weights on the kagome lattice \cite{DFG2020}; Proposition~\ref{prop:dressed} is accordingly presented as a direct consequence of Baxter's equations rather than as an independent result. Along the physical part of Baxter's curve, \eqref{eq:xQexact} increases from $x_1=0$ at $(1:1:0)$, where $\phi\to2\pi/3$, to $x_1=1/2$ at the transition $(1:7:3)$, where $\phi\to0$. The derivation of Proposition~\ref{prop:dressed} relies on the standard dressed-charge analysis of Bethe-ansatz solvable models and is not a rigorous proof; the agreement at the two points is a consistency check of both the proposition and the numerical method.

\section{Results}
\label{sec:results}

\begin{figure*}[t]
\centering
\includegraphics[width=\textwidth]{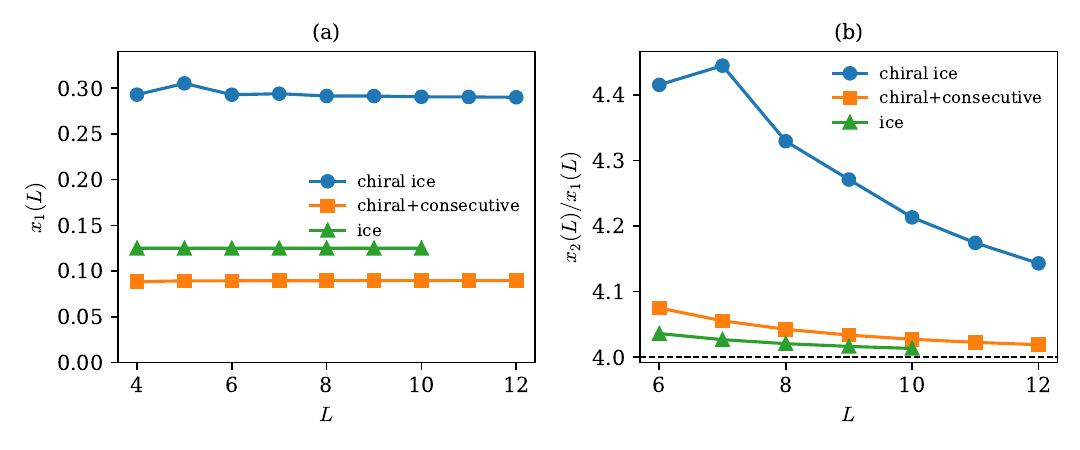}
\caption{(a) Scaling dimension $x_1(L)$ of the lowest state with unit flux, from \eqref{eq:xQ}. (b) Ratio $x_2/x_1$, which tends to $4$ in a Gaussian phase (dashed line).}
\label{fig:sectors}
\end{figure*}
\begin{table*}[t]
\centering
\caption{Central-charge estimates $c(L-2,L)=12[s(L-2)-s(L)]/[\sqrt3\pi((L-2)^{-2}-L^{-2})]$ and flux-sector scaling dimensions \eqref{eq:xQ}.}
\label{tab:sectors}
\begin{tabular}{rccccccccc}
\toprule
 & \multicolumn{3}{c}{ice $(1,1,1)$} & \multicolumn{3}{c}{chiral--consecutive} & \multicolumn{3}{c}{chiral}\\
$L$ & $c$ & $x_1$ & $x_2/x_1$ & $c$ & $x_1$ & $x_2/x_1$ & $c$ & $x_1$ & $x_2/x_1$\\
\midrule
6 & 0.9966 & 0.12492 & 4.036 & 0.9561 & 0.08937 & 4.075 & 1.1174 & 0.29273 & 4.415\\
7 & 0.9969 & 0.12493 & 4.027 & 0.9791 & 0.08943 & 4.056 & 0.9534 & 0.29385 & 4.445\\
8 & 0.9974 & 0.12494 & 4.021 & 0.9948 & 0.08951 & 4.043 & 1.0445 & 0.29139 & 4.329\\
9 & 0.9979 & 0.12495 & 4.017 & 0.9917 & 0.08955 & 4.034 & 0.9670 & 0.29128 & 4.271\\
10 & 0.9983 & 0.12496 & 4.013 & 0.9938 & 0.08959 & 4.027 & 1.0158 & 0.29051 & 4.213\\
11 & --- & --- & --- & 0.9956 & 0.08962 & 4.023 & 0.9824 & 0.29031 & 4.174\\
12 & --- & --- & --- & 0.9962 & 0.08964 & 4.019 & 1.0014 & 0.29000 & 4.143\\
\bottomrule
\end{tabular}
\end{table*}

Table~\ref{tab:data} lists $s(L)$ and the corrected estimates $\tilde s(L)=s(L)-\sqrt3\pi/(12L^2)$ for $4\le L\le12$. At the ice point, computed to $L=10$, the residual $s_\infty-\tilde s(10)=3.2\times10^{-6}$ calibrates the size of the remaining corrections.

\begin{table}[t]
\centering
\caption{Entropy per vertex $s(L)=L^{-1}\ln\Lambda_0$ on cylinders of circumference $L$ and corrected estimates $\tilde s(L)=s(L)-\sqrt3\pi/(12L^2)$.}
\label{tab:data}
\begin{tabular}{rcccc}
\toprule
 & \multicolumn{2}{c}{chiral--consecutive $(1,0,1)$} & \multicolumn{2}{c}{chiral $(0,0,1)$}\\
$L$ & $s(L)$ & $\tilde s(L)$ & $s(L)$ & $\tilde s(L)$\\
\midrule
4 & 0.8574800 & 0.8291394 & 0.5606240 & 0.5322834\\
5 & 0.8477917 & 0.8296538 & 0.5477105 & 0.5295725\\
6 & 0.8424265 & 0.8298307 & 0.5430312 & 0.5304353\\
7 & 0.8390938 & 0.8298397 & 0.5392409 & 0.5299868\\
8 & 0.8369447 & 0.8298595 & 0.5372750 & 0.5301899\\
9 & 0.8354684 & 0.8298702 & 0.5357055 & 0.5301074\\
10 & 0.8344097 & 0.8298752 & 0.5346840 & 0.5301495\\
11 & 0.8336260 & 0.8298785 & 0.5338875 & 0.5301399\\
12 & 0.8330295 & 0.8298805 & 0.5332966 & 0.5301476\\
\midrule
$\infty$ & \multicolumn{2}{c}{$0.829886(1)$} & \multicolumn{2}{c}{$0.53015(1)$}\\
\bottomrule
\end{tabular}
\end{table}

For chiral--consecutive ice, the increments of $\tilde s(L)$ decrease as $10.7$, $5.0$, $3.2$ and $2.0$ (in units of $10^{-6}$) between $L=8$ and $L=12$. The last two increments are both reproduced by a correction $-B/L^4$ with the same coefficient $B=0.102$, which leaves a remaining deficit of $4.9\times10^{-6}$ at $L=12$. The same procedure, applied to the ice point at $L=10$, underestimates the known deficit by $5\%$; including this margin gives $s_\infty=0.829886(1)$. For chiral ice, the even estimates decrease ($0.5301899$, $0.5301495$, $0.5301476$) and the odd ones increase ($0.5301074$, $0.5301399$), so that the two subsequences approach the limit from opposite sides and give $s_\infty=0.53015(1)$. Both values are numerical extrapolations from circumferences $L\le12$, not exact values. The parity oscillation is compatible with the constraint of Proposition~\ref{prop:parity}, which distinguishes even and odd circumferences in chiral ice, although no quantitative link has been established.

Table~\ref{tab:sectors} and Figure~\ref{fig:sectors} collect the central-charge estimates and the scaling dimensions of the flux sectors. At the ice point, $c(L-2,L)$ increases monotonically towards its exact value $1$, which fixes the size of the corrections at these circumferences. For chiral--consecutive ice the estimates follow the same pattern, reaching $c(10,12)=0.9962$. For chiral ice they alternate with the parity of $L$ around $1$ with decreasing amplitude, from $1.117$ and $0.953$ at $L=6,7$ to $1.016$, $0.982$ and $1.001$ at $L=10,11,12$. All three models are therefore consistent with $c=1$.

The flux-sector dimensions distinguish a critical phase from an ordered one. At the ice point $x_1(L)$ varies by less than $5\times10^{-5}$ over $5\le L\le10$, with $x_1(10)=0.12496$, in agreement with the exact value $1/8$ of Proposition~\ref{prop:dressed}. For chiral--consecutive ice $x_1(L)$ increases with decreasing increments towards approximately $0.0897$. For chiral ice $x_1(L)$ does not grow with $L$ but decreases slowly; its even and odd subsequences, extrapolated separately, give $x_1=0.289(1)$. In all three models $x_2/x_1$ decreases monotonically towards $4$, reaching $4.013$, $4.019$ and $4.143$ at the largest circumferences. For chiral ice, where the approach is slowest, the quantity $L^2(x_2/x_1-4)$ takes the values $21.9$, $21.3$, $21.1$ and $20.6$ for $L=9,\dots,12$, so that the deviation from $4$ decreases as $L^{-2}$, as expected for corrections to scaling in a Gaussian phase. The flux spectrum is therefore Gaussian, and no sign of the linear growth of $x_Q(L)$ that an ordered phase would produce is found. The three models differ in the flux dimension $x_1$, which is largest for chiral ice: a unit of flux is costlier in chiral ice than in the full model, and cheapest when the alternating vertices are removed but the consecutive ones are kept.

The entropies can be compared with the independent-vertex estimate $s_{\mathrm P}=\ln(n/8)$, where $n$ is the number of admissible vertices \cite{Pauling1935}. It gives $\ln(5/2)=0.9163$ for the ice point, $4.0\%$ below the exact value; $\ln(9/4)=0.8109$ for chiral--consecutive ice, $2.3\%$ below; and $\ln(3/2)=0.4055$ for chiral ice, $23.5\%$ below. For the frozen consecutive model it gives $\ln(3/4)<0$. The large deficit for chiral ice reflects the constraints along entire lattice lines implied by Propositions~\ref{prop:chiral} and \ref{prop:parity}, which an independent-vertex estimate cannot capture.

\section{The interior of region III}
\label{sec:scan}
\begin{figure}[t]
\centering
\includegraphics[width=\columnwidth]{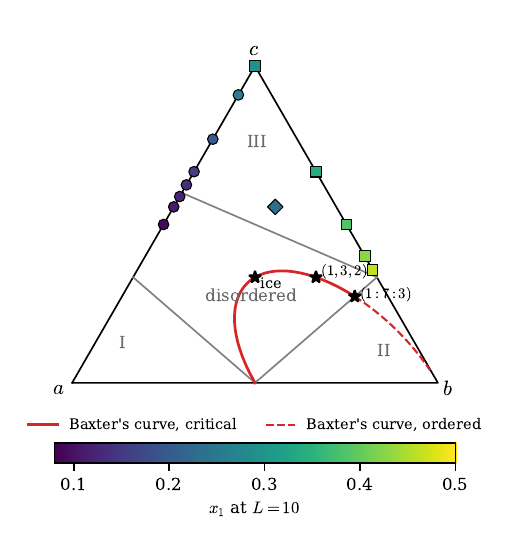}
\caption{Projective plane of weights $(a:b:c)$ in barycentric coordinates. Grey lines: boundaries of the ordered regions I and II and of region III proposed in \cite{Shtengel2005}. Red: Baxter's curve \eqref{eq:baxter}, solid in its critical part and dashed beyond the transition at $(1:7:3)$. Circles: path A; squares: path B; diamond: the interior point $(0.3,0.5,1)$. The colour of each point gives $x_1$ at $L=10$.}
\label{fig:phase}
\end{figure}

\begin{figure}[t]
\centering
\includegraphics[width=\columnwidth]{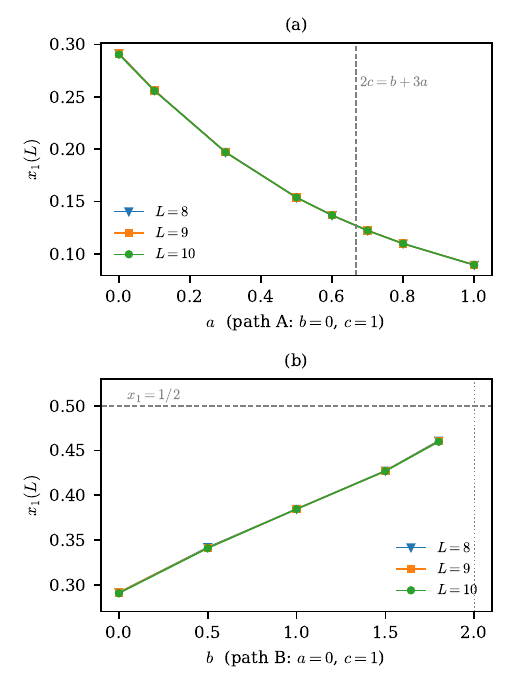}
\caption{Flux dimension $x_1(L)$ at $L=8,9,10$ (a) along path A, $(a,0,1)$, with the proposed boundary of region III at $a=2/3$ (dashed), and (b) along path B, $(0,b,1)$, with the value $x_1=1/2$ of the transition on Baxter's curve (dashed) and the corner $b=2$ of regions II and III (dotted).}
\label{fig:paths}
\end{figure}

The results of Section~\ref{sec:results} concern the extreme point $(0,0,1)$ of region III. To determine whether the region as a whole shares its behaviour, the analysis has been repeated along two paths in the space of weights, with the same transfer matrices, sectors and diagnostics, for $4\le L\le10$. Path A joins chiral--consecutive ice to chiral ice through the points $(a,0,1)$ with $a=1,0.8,0.7,0.6,0.5,0.3,0.1,0$; it enters region III at $a=2/3$, where $2c=b+3a$. Path B runs inside region III through the points $(0,b,1)$ with $b=0.5,1,1.5,1.8$, towards the point $(0,2,1)$ at which the boundaries of regions II and III meet. A further point, $(0.3,0.5,1)$, lies in the interior of region III away from both paths. Figure~\ref{fig:phase} locates all these points in the projective plane of weights, together with the regions of Shtengel and Chayes and Baxter's curve.

Table~\ref{tab:scan} lists the diagnostics at the largest circumference used in the scan, $L=10$. At every point the central-charge estimate lies within $0.016$ of $1$, the flux dimension $x_1(L)$ changes by less than $0.3\%$ between $L=8$ and $L=10$, and the ratio $x_2/x_1$ decreases monotonically towards $4$. No point shows the growth of $x_1(L)$ with $L$ that an ordered phase would produce, nor the collapse of the central-charge estimates that a gap would produce. Every point examined is therefore critical, with $c=1$ and a Gaussian flux spectrum. This conclusion is numerical and rests on circumferences $L\le10$.

\begin{table}[t]
\centering
\caption{Diagnostics along the two paths at $L=10$: central-charge estimate $c(8,10)$, flux dimension $x_1$ and ratio $x_2/x_1$. Points in region III are marked by an asterisk.}
\label{tab:scan}
\begin{tabular}{lcccc}
\toprule
path & $(a,b,c)$ & $c(8,10)$ & $x_1$ & $x_2/x_1$\\
\midrule
A & $(1,0,1)$ & 0.9938 & 0.08959 & 4.027\\
A & $(0.8,0,1)$ & 0.9956 & 0.10992 & 4.028\\
A & $(0.7,0,1)$ & 0.9967 & 0.12243 & 4.029\\
A & $(0.6,0,1)^{*}$ & 0.9980 & 0.13696 & 4.031\\
A & $(0.5,0,1)^{*}$ & 0.9992 & 0.15392 & 4.034\\
A & $(0.3,0,1)^{*}$ & 1.0005 & 0.19716 & 4.051\\
A & $(0.1,0,1)^{*}$ & 1.0039 & 0.25580 & 4.119\\
A, B & $(0,0,1)^{*}$ & 1.0158 & 0.29051 & 4.213\\
B & $(0,0.5,1)^{*}$ & 0.9989 & 0.34086 & 4.099\\
B & $(0,1,1)^{*}$ & 1.0030 & 0.38457 & 4.043\\
B & $(0,1.5,1)^{*}$ & 1.0007 & 0.42703 & 4.066\\
B & $(0,1.8,1)^{*}$ & 0.9991 & 0.45990 & 4.088\\
interior & $(0.3,0.5,1)^{*}$ & 0.9984 & 0.22995 & 4.032\\
\bottomrule
\end{tabular}
\end{table}

Figure~\ref{fig:paths} shows the flux dimension along both paths. Along path A, $x_1$ increases smoothly and monotonically from $0.0896$ at chiral--consecutive ice to $0.2905$ at chiral ice, and the three circumferences $L=8,9,10$ give indistinguishable curves; nothing distinguishes the crossing of the proposed boundary $a=2/3$. Along path B, $x_1$ increases from $0.2905$ to $0.4599$ at $b=1.8$. At the interior point $(0.3,0.5,1)$, $x_1=0.2300$, with $x_1(L)$ varying by less than $0.1\%$ between $L=8$ and $L=10$. The value $1/2$ is the dimension at which, by Proposition~\ref{prop:dressed}, the critical phase ends on Baxter's curve, at the transition $(1:7:3)$ on the boundary of region II. The data along path B are consistent with $x_1$ approaching $1/2$ as $b\to2$, that is, as the boundary of region II is reached, although the circumferences used do not allow this limit to be resolved.

The picture that emerges is that region III, at least along the paths examined, belongs to the same critical phase as the disordered region of Shtengel and Chayes, a single Coulomb phase in which the flux dimension varies continuously with the weights. The proposed boundary $2c=b+3a$ marks the limit of positive switch weights, not a phase transition.

\section{Conclusions}
\label{sec:conclusions}

Two restrictions of the symmetric twenty-vertex model on the triangular lattice have been examined: chiral ice, the extreme point $(0,0,1)$ of the region of the phase diagram dominated by chiral vertices, and chiral--consecutive ice, $(1,0,1)$, which excludes only the alternating vertices. Both points lie outside every exactly solvable family known for this model. For Baxter's Bethe-ansatz curve the exclusion follows from the identity $b-c=(a-c)^2/a$, which forbids $b<c$. Kelland's surface reduces on the symmetric slice to Baxter's curve; chiral--consecutive ice violates Kelland's one-magnon commutation conditions, chiral ice cannot be reached at finite values of his parameters, both points violate his unstudied branch, and a computer-algebra certificate shows that neither belongs to the Zariski closure of the surface. A single three-term Pl\"ucker relation excludes both points from the free-fermionic subvariety, including its column rescalings. Chiral ice has been characterised as the ice model in which every vertex has exactly one through line, a property that imposes a parity constraint on every closed lattice line. The residual entropies have been computed with transfer matrices on cylinders of circumference up to twelve sites and extrapolated with the conformal correction $\sqrt3\pi c/(12L^2)$, whose amplitude has been derived for the triangular geometry and verified at two exactly solved points, one of which is also the symmetric free-fermionic point. The results are $s=0.829886(1)$ for chiral--consecutive ice and $s=0.53015(1)$ for chiral ice. Both models have central charge $c=1$ and a Gaussian spectrum of flux sectors, with scaling dimensions $x_1\simeq0.0897$ and $x_1=0.289(1)$ respectively, against the exact value $x_1=1/8$ at the ice point, which has been derived from Baxter's Bethe ansatz together with the closed form $x_Q=(2\pi-3\phi)Q^2/(4\pi)$ along his curve. Chiral ice is therefore critical, and the extreme point of the region in which ordering by disorder had been suggested belongs to a Coulomb phase. A scan along two paths through the region, one of which crosses its proposed boundary, finds every point critical with $c=1$ and a Gaussian flux spectrum, and a flux dimension that varies continuously from $0.0896$ at chiral--consecutive ice to $0.29$ at chiral ice and towards $1/2$ as the boundary of the antiferroelectric region is approached. Region III therefore appears to belong, along the paths examined, to the same Coulomb phase as the disordered region, its proposed boundary marking the limit of positive switch weights rather than a phase transition. Whether the flux dimension reaches exactly $1/2$ on the boundary of region II, as it does on Baxter's curve, is left as a question for larger circumferences or an analytic treatment.

\begin{acknowledgments}
J.B. appreciates fruitful discussions with Maria Vallespir-Socias, Joana Rossell\'o, Margalida Batle, Maria del Mar Batle, and Regina Batle. J.B. thanks Peichang Ouyang for a question on his work on quantum error-correcting codes from aperiodic tilings that led to the present study. J.B. received no funding for the present work.

\end{acknowledgments}

\section*{Data availability}
The algorithms used in this work are given in pseudocode in Appendix~\ref{app:algo}. The data that support the findings of this study are available from the author on reasonable request.

\appendix
\section*{Appendix: Algorithms}
\label{app:algo}

The three computational procedures of this work are given below. Algorithm~\ref{alg:closure} is the elimination behind the closure statement of Theorem~\ref{thm:kelland}. The row operator is defined by \eqref{eq:Tdef}; Algorithm~\ref{alg:T} gives the corresponding matrix--vector multiplication used in the numerical calculations, with the bonds labelled as in \eqref{eq:word}. Algorithm~\ref{alg:sectors} summarises the sector computation and the finite-size analysis.

\begingroup\small\noindent\hrule\smallskip
\refstepcounter{algo}\label{alg:closure}\noindent\textbf{Algorithm~\thealgo}\quad Zariski-closure certificate for Kelland's surface (Theorem~\ref{thm:kelland})
\smallskip\hrule\smallskip
\begin{algorithmic}[1]
\Require cleared equations of \eqref{eq:k19abc}--\eqref{eq:k19e} in the variables $u_1,\dots,u_{10},I$
\State $J\gets$ ideal generated by $u_5-u_6,\ u_7-u_8,\ u_9-u_{10}$, $Iu_7-(u_2u_3u_7+u_1u_5u_9)$, $Iu_9-(u_3u_1u_9+u_2u_7u_5)$, $Iu_5-(u_1u_2u_5+u_3u_7u_9)$, and $u_1u_2u_3u_4-[I^2-(u_2u_3+u_1u_3+u_1u_2)I+u_1u_2u_3(u_1+u_2+u_3)]$
\State $J_{\mathrm{sat}}\gets J:(u_1u_2u_3u_5u_7u_9)^\infty$ \Comment{removes the components where Kelland's divisions fail}
\State $G\gets$ Gr\"obner basis of $J_{\mathrm{sat}}\cap\mathbb Q[u_1,\dots,u_{10}]$ \Comment{elimination of $I$}
\For{$p\in\{(1,1,1),(1,3,2),(1,\tfrac79,\tfrac13),(1,0,1),(0,0,1)\}$, with $u_{1,2,3}=a$, $u_4=b$, $u_{5,\dots,10}=c$}
\State report whether $g(p)=0$ for every $g\in G$
\EndFor
\Ensure all generators vanish at the three points of Baxter's curve and not all vanish at $(1,0,1)$ and $(0,0,1)$
\end{algorithmic}
\smallskip\hrule\par\endgroup\medskip

\begingroup\small\noindent\hrule\smallskip
\refstepcounter{algo}\label{alg:T}\noindent\textbf{Algorithm~\thealgo}\quad Application of the row-to-row transfer matrix on a cylinder of circumference $L$
\smallskip\hrule\smallskip
\begin{algorithmic}[1]
\Require vector $v$ indexed by the state $\sigma=(u_x,v_x)_{x=1}^{L}$ of a cut; vertex tensor $A[u,v,h;u',\tilde v,h']=w(h',u',\tilde v,1-h,1-u,1-v)$, cf.\ \eqref{eq:word}
\State $w\gets v\otimes(\lvert00\rangle+\lvert11\rangle)$ on two auxiliary bits $(h,h_0)$ \Comment{$h$ holds $h_0$; $h_0$ keeps a copy}
\For{$x=1,\dots,L$}
\State contract the bits $(u_x,v_x,h)$ of $w$ with the input indices of $A$ and replace them by the output bits $(u'_x,\tilde v_x,h)$ \Comment{$\tilde v_x=v'_{x-1}$, and $h$ now holds $h_x$}
\EndFor
\State $u\gets\sum_{h}w[\dots,h,h_0=h]$ \Comment{closure $h_L=h_0$ of the ring of horizontal bonds}
\State relabel $\tilde v_x$ as $v'_{x-1}$
\Ensure $u=Tv$
\end{algorithmic}
\smallskip\hrule\par\endgroup\medskip

\begingroup\small\noindent\hrule\smallskip
\refstepcounter{algo}\label{alg:sectors}\noindent\textbf{Algorithm~\thealgo}\quad Sector eigenvalues and finite-size analysis
\smallskip\hrule\smallskip
\begin{algorithmic}[1]
\For{$L=4,\dots,L_{\max}$}
\For{$Q\in\{0,1,2\}$}
\State $S_Q\gets\{\text{states with }N_\uparrow=L+Q\}$
\State $\Lambda_Q\gets$ leading eigenvalue of $T$ restricted to $S_Q$, by the implicitly restarted Arnoldi method with Algorithm~\ref{alg:T}
\EndFor
\State $s(L)\gets L^{-1}\ln\Lambda_0$; \ $\tilde s(L)\gets s(L)-\sqrt3\pi/(12L^2)$
\State $c(L-2,L)\gets12[s(L-2)-s(L)]/[\sqrt3\pi((L-2)^{-2}-L^{-2})]$; \ $x_Q(L)\gets L\ln(\Lambda_0/\Lambda_Q)/(\sqrt3\pi)$
\EndFor
\Ensure checks: $\operatorname{Tr}T^M$ equals a direct enumeration of all $2^{3LM}$ orientations for $(L,M)\in\{(2,2),(2,3),(3,2)\}$; Algorithm~\ref{alg:T} agrees with an independent tensor contraction; $T$ applied to a vector of one sector has no component in any other; the leading eigenvalue on the full space equals $\Lambda_0$
\end{algorithmic}
\smallskip\hrule\par\endgroup

\end{document}